\documentclass[conference]{IEEEtran}
\IEEEoverridecommandlockouts

\usepackage{cite}
\usepackage{amsmath,amssymb,amsfonts}
\usepackage{algorithmic}
\usepackage{graphicx}
\usepackage{textcomp}
\usepackage{subcaption}
\usepackage{xcolor}
\usepackage{url}
\def\BibTeX{{\rm B\kern-.05em{\sc i\kern-.025em b}\kern-.08em
    T\kern-.1667em\lower.7ex\hbox{E}\kern-.125emX}}
\begin{document}

\title{LatentSpeech: Latent Diffusion for Text-To-Speech Generation\\
}

\author{\IEEEauthorblockN{Haowei Lou}
\IEEEauthorblockA{
\textit{UNSW Sydney}\\
Kensington, Australia \\
0009-0009-1359-872X}
\and
\IEEEauthorblockN{Helen Paik}
\IEEEauthorblockA{
\textit{UNSW Sydney}\\
Kensington, Australia \\
0000-0003-4425-7388}
\and
\IEEEauthorblockN{Pari Delir Haghighi}
\IEEEauthorblockA{
\textit{Monash University}\\
Clayton, Australia \\
0000-0001-9922-1214}
\and
\IEEEauthorblockN{Wen Hu}
\IEEEauthorblockA{
\textit{UNSW Sydney}\\
Kensington, Australia \\
0000-0002-4076-1811}
\and
\IEEEauthorblockN{Lina Yao}
\IEEEauthorblockA{
\textit{UNSW Sydney}\\
Kensington, Australia \\
0000-0002-4149-839X}
}

\maketitle
\begin{abstract}
Diffusion-based Generative AI gains significant attention for its superior performance over other generative techniques like Generative Adversarial Networks and Variational Autoencoders. While it has achieved notable advancements in fields such as computer vision and natural language processing, their application in speech generation remains under-explored. Mainstream Text-to-Speech systems primarily map outputs to Mel-Spectrograms in the spectral space, leading to high computational loads due to the sparsity of MelSpecs. To address these limitations, we propose LatentSpeech, a novel TTS generation approach utilizing latent diffusion models. By using latent embeddings as the intermediate representation, LatentSpeech reduces the target dimension to 5\% of what is required for MelSpecs, simplifying the processing for the TTS encoder and vocoder and enabling efficient high-quality speech generation. This study marks the first integration of latent diffusion models in TTS, enhancing the accuracy and naturalness of generated speech. Experimental results on benchmark datasets demonstrate that LatentSpeech achieves a 25\% improvement in Word Error Rate and a 24\% improvement in Mel Cepstral Distortion compared to existing models, with further improvements rising to 49.5\% and 26\%, respectively, with additional training data. These findings highlight the potential of LatentSpeech to advance the state-of-the-art in TTS technology
\end{abstract}

\begin{IEEEkeywords}
Text-to-Speech, Speech Synthesis, Latent Diffusion, Generative Artificial Intelligence
\end{IEEEkeywords}

\section{Introduction}
Generative AI has made significant strides in recent years. It revolutionises various fields with its ability to generate high-quality data. Among numerous GAI techniques, diffusion-based generative models have garnered increased attention for their superior performance compared to other methods such as Generative Adversarial Networks~\cite{goodfellow2020generative} and Variational Autoencoders~\cite{kingma2013auto}. Diffusion models demonstrate remarkable advancements in areas like image generation~\cite{rombach2022high}, large language models~\cite{ramesh2022hierarchical}, and video generation~\cite{ho2022imagen}. 

Mainstream Text-to-Speech (TTS) systems, which convert linguistic context to speech using deep learning approaches, have explored the application of advanced deep learning techniques in speech generation. For instance, Tacotron~\cite{wang2017tacotron} employs a sequence-to-sequence framework for speech generation, FastSpeech~\cite{ren2019fastspeech} uses a transformer architecture to enable parallel computation and address issues like word skipping, and StyleSpeech~\cite{lou2024stylespeechparameterefficientfinetuning} enhances phoneme and style embedding efficiency to improve speech quality.

One challenge for mainstream TTS methods is their reliance on MelSpec as an intermediate representation. 
MelSpecs are characterized by high sparsity, which leads to significant computational and parameter demands to process the sparse content. Each MelSpec represents the frequency content of a speech over time, resulting in a large and mostly empty matrix where only a few values carry significant information. This sparsity requires models to allocate extensive computational resources and memory to process and store these large matrices.

There are methods that attempt to generate MelSpecs using diffusion models \cite{zhang2023survey}, and approaches like DiffVoice \cite{liu2023diffvoice} that employ latent diffusion with MelSpecs as an intermediate representation. Some approaches, such as FastSpeech 2~\cite{ren2020fastspeech}, have explored direct speech generation without relying on MelSpec. The potential of using latent embeddings directly in the audio space as the intermediate representation for TTS systems remains underexplored. 

In this study, we propose LatentSpeech, a novel diffusion-based TTS framework that operates in the latent space. Our method leverages the advantages of diffusion methods in capturing intricate details in latent embeddings. It results in a more effective learning process, thereby enhancing the quality of generated speech. The main contributions are:


\begin{figure}[t]
    \centering
    \begin{subfigure}{0.8\linewidth}
        \centering                
        \includegraphics[width=\linewidth]{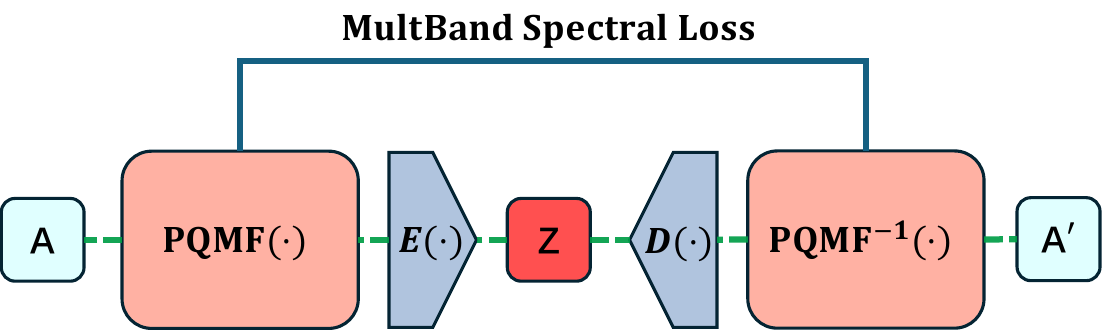}
        \label{fig:ae}
    \end{subfigure}
    \begin{subfigure}{0.9\linewidth}
        \centering
        \includegraphics[width=\linewidth]{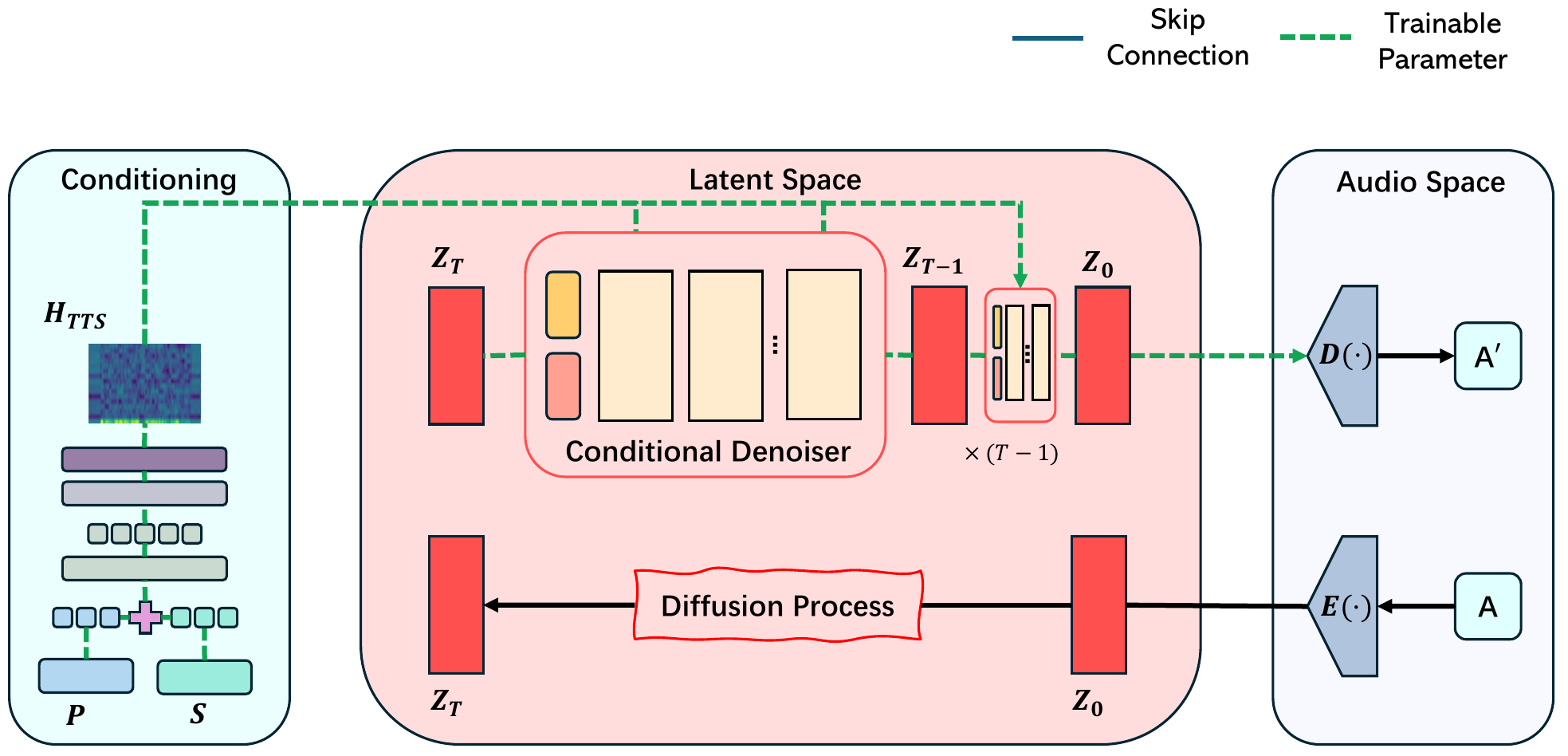}
    \end{subfigure}
    \caption{LatentSpeech}
    \label{fig:latent_speech}
\end{figure}
\begin{enumerate}
    \item LatentSpeech is the first approach to leverage latent diffusion in TTS for directly generating high-quality speech in the audio space. Unlike other methods that apply latent diffusion on Mel-Spectrogram, LatentSpeech applies it directly on raw audio.
    \item LatentSpeech reduces the intermediate representation dimension to 5\% of MelSpecs by using latent embeddings. This reduction simplifies the processing for the TTS encoder and vocoder and enables efficient high-quality speech generation.
    \item LatentSpeech achieves a 25\% improvement in Word Error Rate and a 24\% improvement in Mel Cepstral Distortion, with improvements rising to 49.5\% and 26\%, respectively, with more training data.
\end{enumerate}

\begin{figure}
    \centering
    \begin{subfigure}[b]{0.9\linewidth}
        \centering
        \begin{subfigure}{\textwidth}
            \centering
            \begin{subfigure}{0.4\textwidth}
                \centering
                \includegraphics[width=\linewidth]{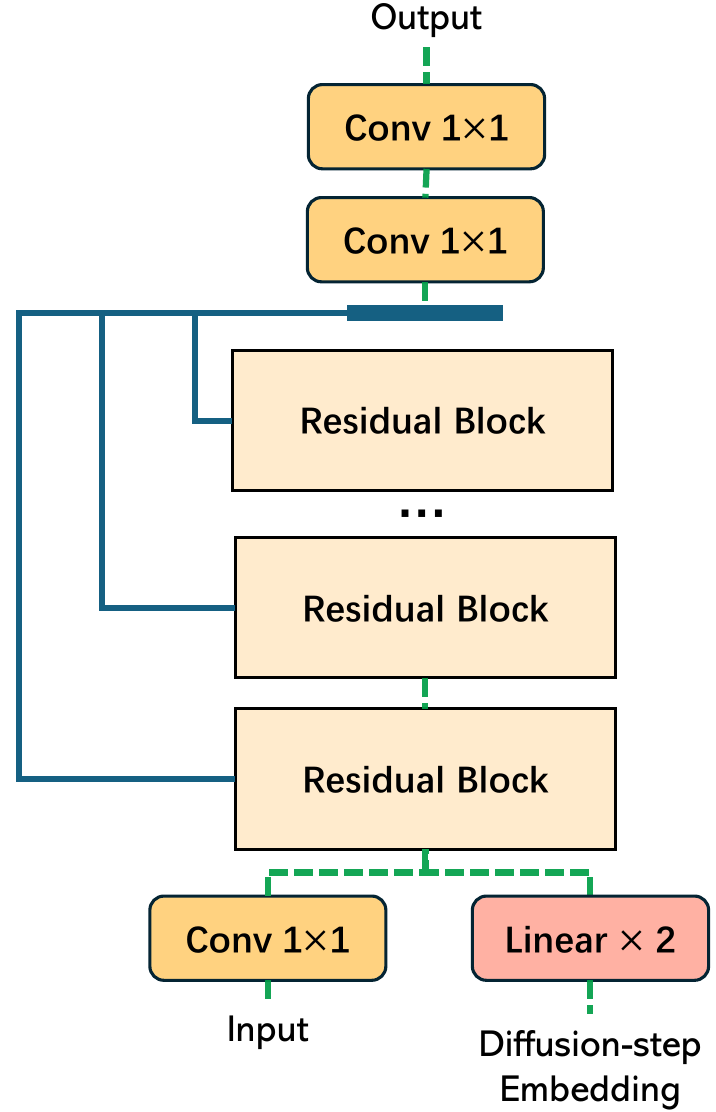}
                \caption{Conditional Denoiser}
                \label{fig:conditional_denoiser}
            \end{subfigure}   
            \begin{subfigure}{0.53\linewidth}
                \centering
                \includegraphics[width=\linewidth]{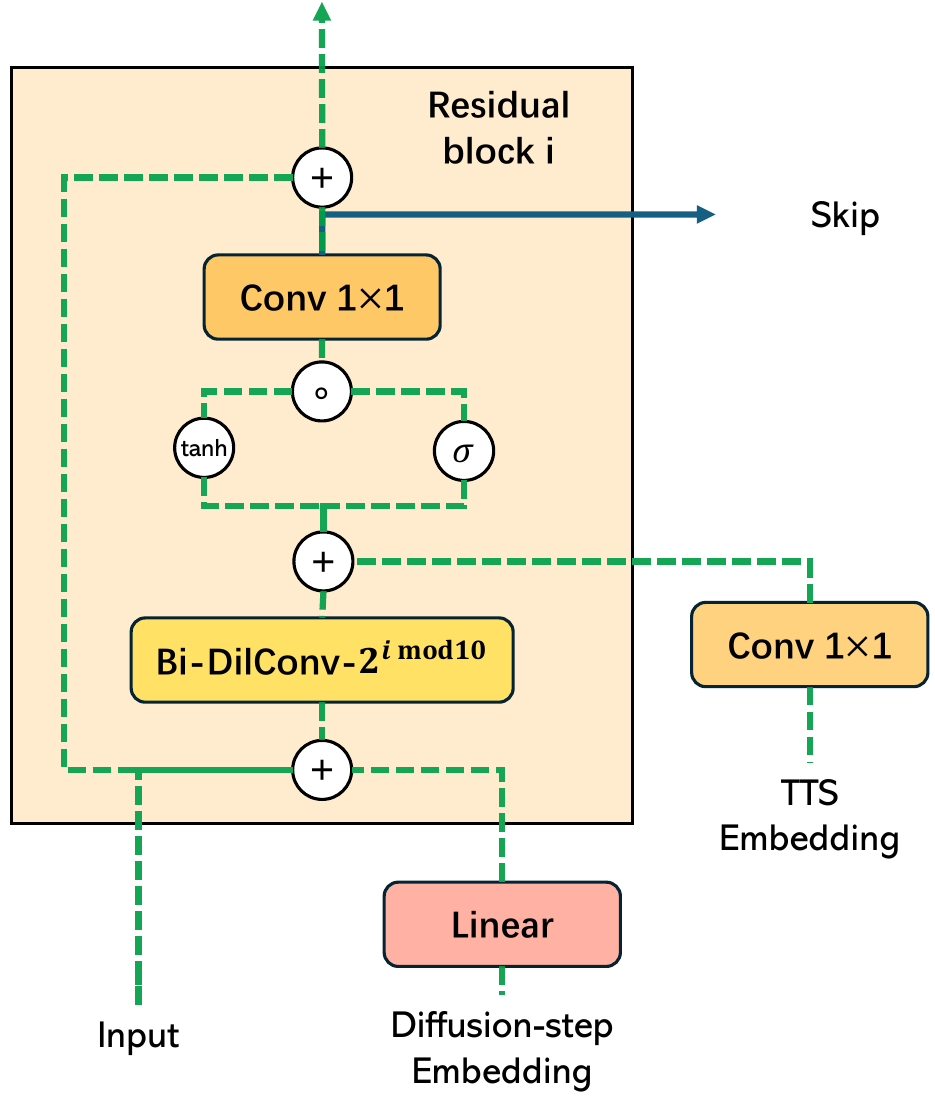}
                \caption{Residual Block}
                \label{fig:residual_block}
            \end{subfigure}
        \end{subfigure}  
    \end{subfigure}

    \caption{Conditional Denoiser Diagram}
\end{figure}

\section{LatentSpeech}\label{sec:method}
In this section, we introduce the architecture of LatentSpeech. We first encode speech~$A$ into latent space using an Autoencoder~(AE). Then, we set latent embeddings as the intermediate representation~$Z$ and train a diffusion-based TTS model to map embeddings. In the end, we generate speech directly from the latent space to the audio space using the trained decoder. An overview of the entire system is provided in Figure~\ref{fig:latent_speech}.

\subsection{Latent Encoder}\label{sec:lfe}
To lower the computation demand of training TTS system and sparsity of intermediate representation. 
We follow a similar training setup to RAVE~\cite{caillon2021rave} to train an Autoencoder to encode speech from audio space to latent space. Specifically, given a raw waveform~$A \in \mathbb{R}^{L_{audio}}$ where $L_{audio}$ is the number of time points in the speech. We first apply a multi-band decomposition to the raw speech using Pseudo Quadrature Mirror Filters (PQMF)~\cite{nguyen1994near}.
\begin{equation}
\mathbf{PQMF}(A) = \mathbb{R}^{N \times L_{sub}}, \, L_{audio}=N \times L_{sub}
\end{equation}
$N$ is the number of frequency sub-bands and $L_{sub}$ is the number of time points in each sub-band.
An encoder is applied~$\mathbf{E}(\cdot)$ to encode $\mathbf{PQMF}(A)$ into latent space~$Z \in \mathbb{R}^{N \times L_{\text{latent}}}$. Here, $N$ denotes the number of channels $L_{\text{latent}}$ represents the latent space temporal resolution. 
The latent embeddings are passed into a decoder~$\mathbf{D}(\cdot)$ to reconstruct $\mathbf{PQMF}(A)$, yielding $\mathbf{D}(Z)$. The resultant multi-band speech is then processed using the inverse PQMF function to produce the reconstructed speech, \(A' = \mathbf{PQMF}^{-1}(\mathbf{D}(Z))\). 
We use the multiscale spectral distance in the multi-band speech as the loss function~\cite{engel2020ddsp} to train the encoder and decoder. $N$ and $L$ will be used in the following sections to denote the number of channels and time resolution in the latent space.

\subsection{Text-to-Speech Encoder}\label{sec:tts_encode}
TTS encoder transforms linguistic inputs to TTS embedding, which serves as conditions for the diffusion model to map latent embedding. In this work, we adopt the transformer-based TTS system StyleSpeech~\cite{lou2024stylespeechparameterefficientfinetuning} as our TTS encoder. It includes the following key components: an acoustic pattern encoder, a duration adapter, and an integration encoder, each consisting of multiple layers of Feed-Forward Transformers (FFT Blocks)~\cite{ren2019fastspeech}.

Given sequences of phonemes~\(P\) and styles~\(S\) linguistic input, the Acoustic Pattern Encoder (APE) transforms input text into sequences of phoneme \(H_P = (h_{P1}, \ldots, h_{Pn})\) and style embeddings \(H_S = (h_{S1}, \ldots, h_{Sn})\). The phoneme and style embeddings are fused to produce acoustic embedding \(H = H_P + H_S\).
The Duration Adapter controls the duration of acoustic embeddings to align acoustic embedding with real speech. It has two main components: the \textit{duration predictor} and the \textit{length regulator}. The duration predictor estimates the duration of each acoustic feature~\(L = \{l_1, \ldots, l_n\}\), where \(m = \sum_{i=0}^{N} l_i\). These durations adjust the length of each acoustic embedding to the adaptive embedding~\(H_L = \{h_{l1}, \ldots, h_{lm}\}\).
The adaptive embeddings~\(H_L\) are then passed through the embedding generator to generate the TTS embedding~\(H_{TTS}\). This is followed by a linear layer to broadcast~\(H_{TTS}\) to the dimensions of the latent embedding~\(Z\).

\subsection{Latent Diffusion}\label{sec:ld}
Diffusion model is a probabilistic generative model that learns to produce data that match latent embedding distribution~$p(Z)$, by denoising a normally distributed variable through a reverse Markov Chain of length~$T$. 
We define $q(Z_0)$ as the data distribution of the latent embedding $Z \in \mathbb{R}^{N \times L}$. Let $Z_t \in \mathbb{R}^{N \times L}$ for $t = 0, 1, \ldots, T$ represent the forward diffusion process:

\begin{equation}
    q(Z_{1:T} | Z_0) = \prod_{t=1}^{T} q(Z_t | Z_{t-1})
\end{equation}
where Gaussian noise~$\mathcal{N}(\cdot)$ is gradually added to the Markov chain from $Z_0$ to $Z_T$ until $q(Z_T) \sim \mathcal{N}(0,I)$. 
\begin{equation}
    q(Z_t | Z_{t-1}) = \mathcal{N}(Z_t;\sqrt{\alpha_t}Z_{t-1}, (1-\alpha_t)\mathbf{I})
\end{equation}
Here, $\alpha_t$ refers to the scaling factor that controls the amount of noise added at diffusion step $t$.
Then, we apply a conditional denoiser~$P_\theta(\cdot)$ parameterized by~$\theta$ to reverse the diffusion process and gradually reconstruct the original latent embeddings~$Z_0$ from the noisy latent embeddings~$Z_T$, as illustrated in Figure~\ref{fig:conditional_denoiser}. 
\begin{equation}
    p_{\theta}(Z_0 | Z_{T:1}) = \prod_{t=1}^{T} p_\theta(Z_{t-1}|Z_{t},t_{embed},H_{TTS})
\end{equation}
Specifically, we apply a 128-dimensional positional encoding~\cite{vaswani2017attention} at diffusion step~$t$ to represent the diffusion-step embedding~$t_{embed}$. 
$t_{embed}$ is broadcasted to the $L$-dimension, $t_{embed} \in \mathbb{R}^{L}$, to match the temporal resolution of latent embedding~$Z$. The TTS embedding~$H_{TTS}$, obtained in Section~\ref{sec:tts_encode}, serves as a conditional input for the denoiser to guide the reverse diffusion process. The dimension of $H_{TTS}$ is as same as the latent embedding dimension~$Z \in \mathbb{R}^{N \times L}$.
The denoiser is constructed using several layers of residual blocks built with bidirectional dilated convolution kernels, similar to those applied in diffusion-based neural vocoders \cite{kong2020diffwave}. More details on the architecture of the denoiser and the residual blocks can be found in Figures~\ref{fig:conditional_denoiser} and \ref{fig:residual_block} respectively.

\textbf{Training}: In the training stage, we define the transition probability $p_\theta(Z_{t-1}|Z_{t})$ is parameterized as 
\begin{equation}
\mathcal{N}(Z_{t-1};\mu_{\theta}(Z_{t},t), \sigma_{\theta}(Z_{t},t)^2I) 
\end{equation}
$\mu_\theta$ is the mean embedding and $\sigma_\theta$ is a real number as the standard deviation.
We follow a closed-form diffusion model calculation method proposed in \cite{ho2020denoising} to accelerate computation and avoid Monte Carlo estimates. Specifically, we first define the variance schedule~${\beta}_{t=1}^{T}$:
\begin{align}
    \alpha_t = 1 - \beta_t, \quad \hat{\alpha}_{t} = \prod_{s=1}^{t}\alpha_s, \quad \\
    \hat{\beta_{t}} = \frac{1-\hat{\alpha}_{t-1}}{1-\hat{\alpha}_{t}}\beta_{t}, \quad t > 1, \quad \hat{\beta_{1}} = \beta_1
\end{align}
Then, the parameterizations of $\mu_\theta$ and $\sigma_\theta$ are defined by:
\begin{equation}
    \mu_{\theta}(\cdot) = \frac{1}{\sqrt{\alpha_t}}(Z_t - \frac{\beta_t}{\sqrt{1-\hat{\alpha}}}f_{\theta}(Z_{t},t,H_{TTS})),\, \sigma_{\theta}(\cdot) = \sqrt{\beta_{t}}
\end{equation}
Here, $f_{\theta}(Z_{t},t,H_{TTS})$ is our proposed conditional denoiser r, which takes the diffusion step embedding~$t_{embed}$ and TTS embedding~$H_{TTS}$ as conditional inputs to predict the noise $\epsilon_t$ added in the forward diffusion process at step $t$. The training objective is to optimize the parameters to reduce the following loss function:
\begin{equation}
    L = \mathbb{E}_{Z_0,\epsilon,t,H_{TTS}} \left\| \epsilon - f_\theta(Z_t, t, H_{TTS}) \right\|^2 
\end{equation}

\textbf{Inference}: In the inference stage, we sample $Z_T \sim \mathcal{N}(0,I)$. We use the trained denoiser~$f_{\theta}({\cdot})$ predicts the noise~$\epsilon_{t}$ added to the latent embeddings at $t$ for $t=T,T-1,\dots,1$. This noise is iteratively subtracted from~$Z_T$ until the latent embedding~$Z_0$ is reconstructed.
\begin{equation}
    \epsilon_{t} = f_{\theta}(Z_{t-1},t_{embed},H_{TTS})
\end{equation}

\subsection{Vocoder}
The trained decoder $\mathbf{D}(\cdot)$ described in Section~\ref{sec:lfe} serves as a vocoder to reconstruct speech using the latent embeddings produced by the diffusion denoising process outlined in Section~\ref{sec:ld}. Specifically, the denoised latent embeddings \(Z \in \mathbb{R}^{N \times T}\) are input into the decoder $\mathbf{D}(\cdot)$. The decoder converts these features back into multi-band speech, which is then processed using the inverse PQMF function, $\mathbf{PQMF}^{-1}(\cdot)$. This function combines the sub-band speech signals back into a single speech waveform to generate the final reconstructed speech signal \(A'\).

\begin{figure}[t]
    \centering
    \begin{subfigure}[t]{0.12\textwidth}
        \includegraphics[width=\textwidth]{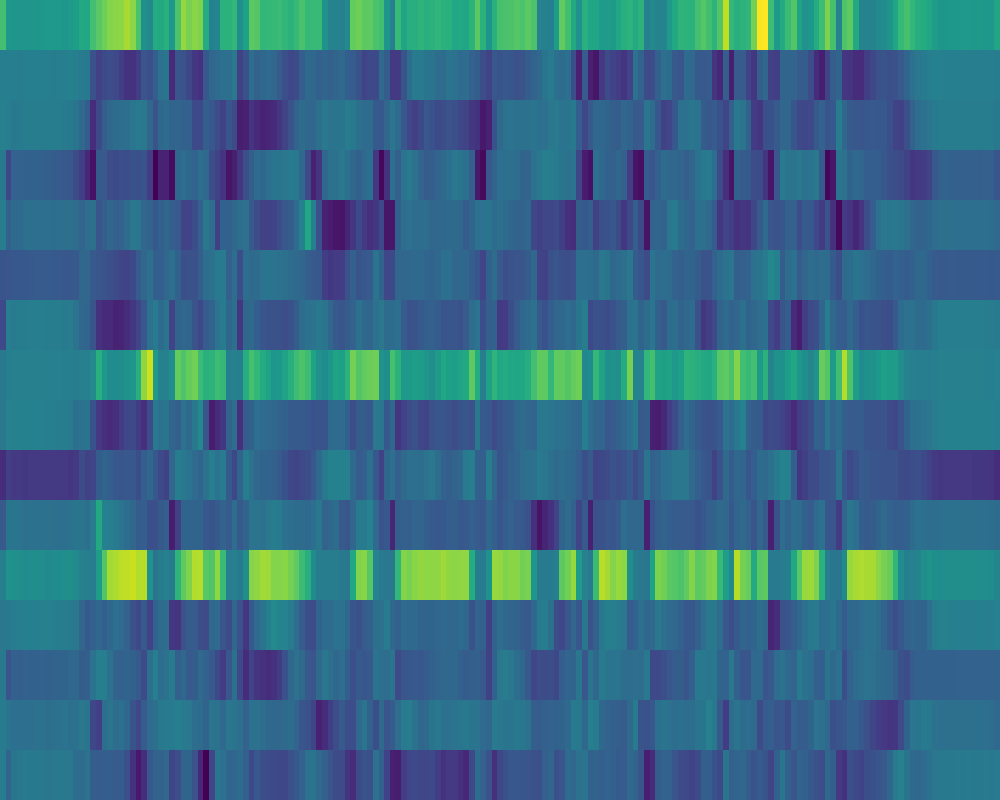}
        \caption{TTS Embed}
        \label{fig:tts_embed}
    \end{subfigure}
    \begin{subfigure}[t]{0.12\textwidth}
        \includegraphics[width=\textwidth]{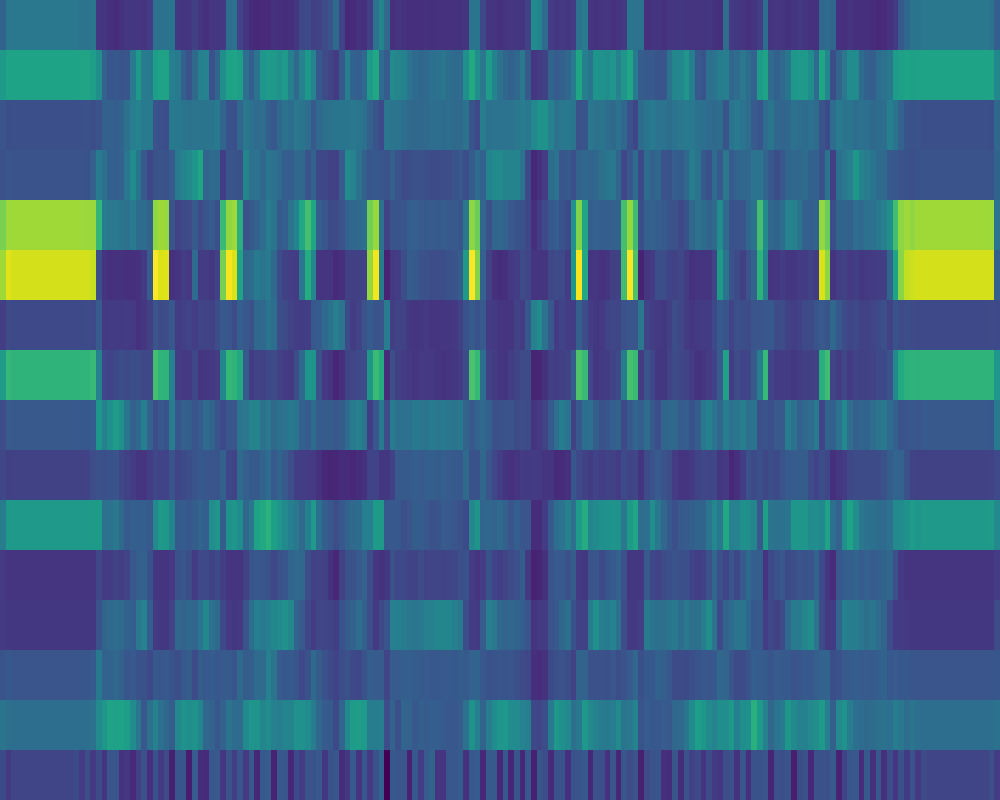}
        \caption{Latent Embed}
        \label{fig:latent_real}
    \end{subfigure}
    \begin{subfigure}[t]{0.12\textwidth}
        \includegraphics[width=\textwidth]{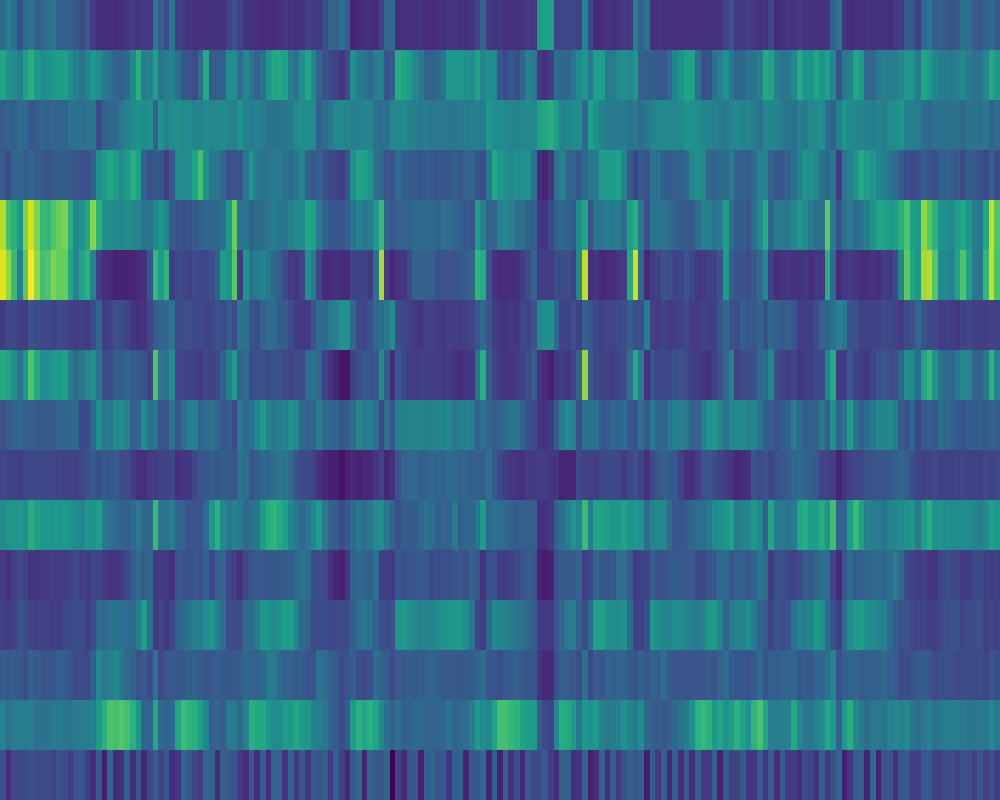}
        \caption{Generate Embed}
        \label{fig:latent_fake}
    \end{subfigure}
    \begin{subfigure}[t]{0.12\textwidth}
        \includegraphics[width=\textwidth]{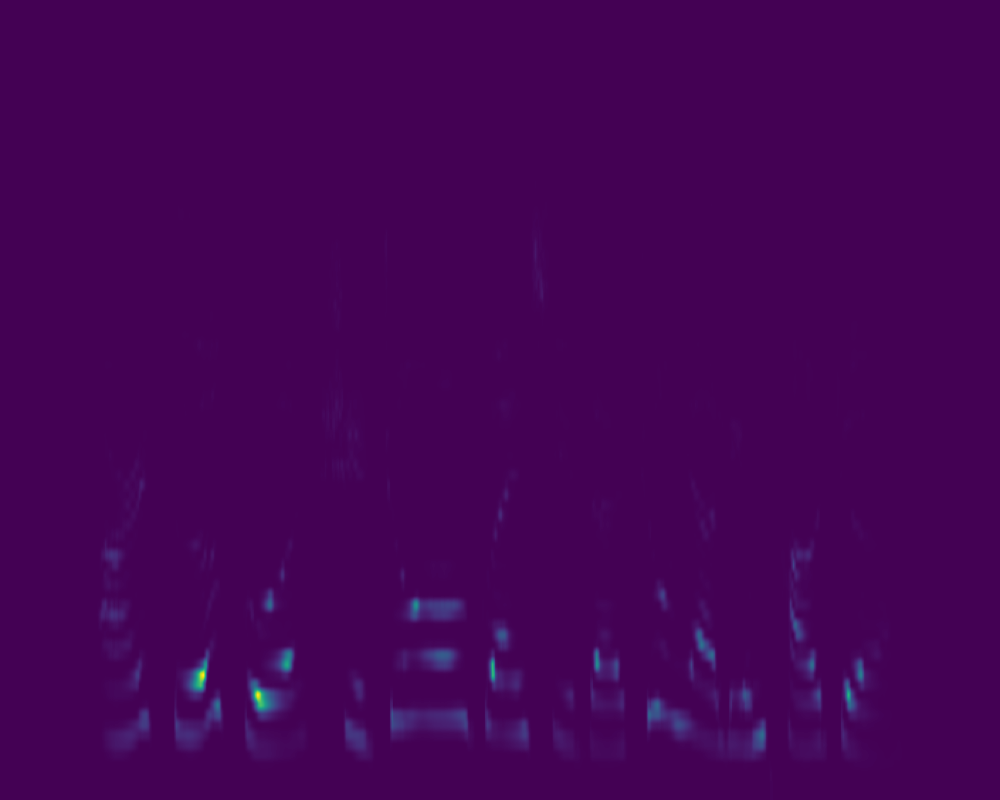}
        \caption{MelSpec}
        \label{fig:real_mel}
    \end{subfigure}
    \begin{subfigure}[t]{0.12\textwidth}
        \includegraphics[width=\textwidth]{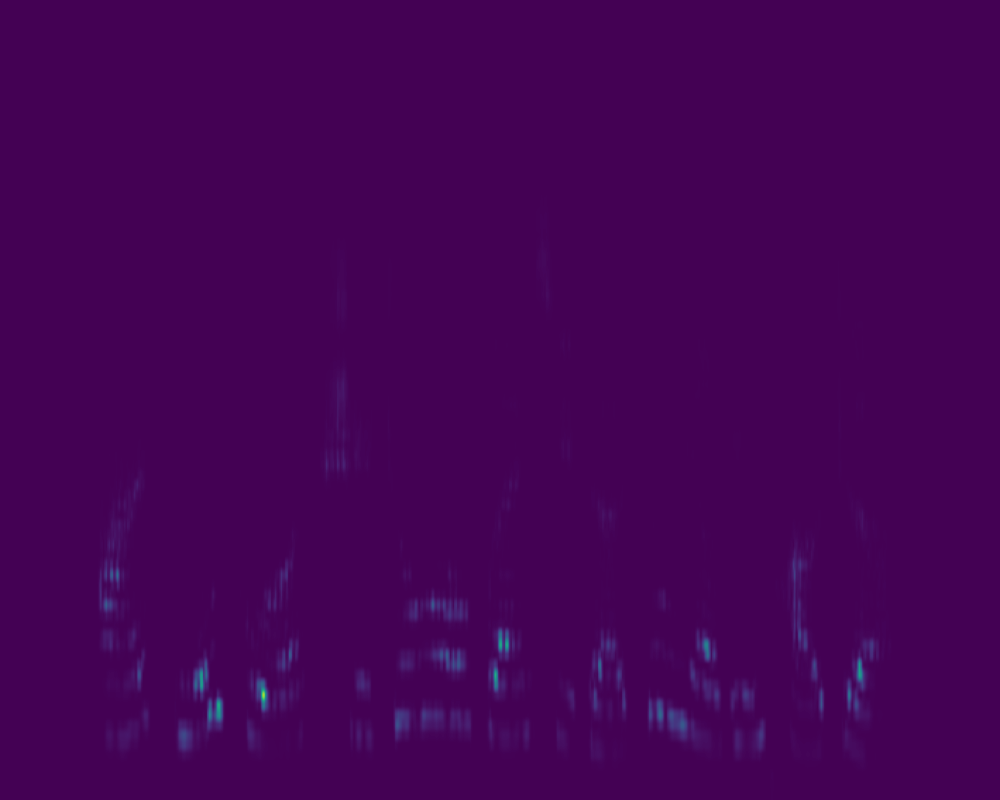}
        \caption{Generate MelSpec}
        \label{fig:fake_mel}
    \end{subfigure}
    \caption{Embed Visualization}
    \label{fig:audio_visual}
\end{figure}

\begin{table*}[!t]
\centering

\begin{tabular}{lccccc}
\hline
\textbf{Model} & \textbf{WER~(\(\downarrow\))} & \textbf{WER-P~(\(\downarrow\))} & \textbf{WER-S~(\(\downarrow\))} & \textbf{MCD~(\(\downarrow\))} & \textbf{PESQ~(\(\uparrow\))} \\
\hline
FastSpeech & 0.419 $\pm$ 0.184 & 0.211 $\pm$ 0.128 & 0.342 $\pm$ 0.153 & 13.003 $\pm$ 4.081 & 1.054 $\pm$ 0.055 \\
StyleSpeech & 0.312 $\pm$ 0.156 & 0.220 $\pm$ 0.140 & 0.171 $\pm$ 0.112 & 12.843 $\pm$ 4.009 & 1.058 $\pm$ 0.059 \\
\hline
4k Sentence \\
\hline
LatentSpeech (w l) & 0.275 $\pm$ 0.150 & 0.184 $\pm$ 0.136 & 0.177 $\pm$ 0.105 & \textbf{9.723 $\pm$ 1.791} & \textbf{1.055 $\pm$ 0.014} \\
LatentSpeech (w/o l) & \textbf{0.235 $\pm$ 0.136} & \textbf{0.142 $\pm$ 0.119} & \textbf{0.151 $\pm$ 0.099} & 15.724 $\pm$ 2.479 & 1.047 $\pm$ 0.019 \\
\hline
9k Sentence \\
\hline
LatentSpeech (w l) & \textbf{0.153 $\pm$ 0.106} & \textbf{0.075 $\pm$ 0.083} & \textbf{0.112 $\pm$ 0.082} & \textbf{9.498 $\pm$ 1.620} & 1.058 $\pm$ 0.014 \\
LatentSpeech (w/o l) & 0.168 $\pm$ 0.118 & 0.087 $\pm$ 0.099 & 0.122 $\pm$ 0.091 & 15.080 $\pm$ 2.276 & \textbf{1.063 $\pm$ 0.038} \\
\hline
\end{tabular}%

\caption{Evaluation Results of TTS systems. \textbf{(\(\downarrow\))} indicates that lower values are better, and \textbf{(\(\uparrow\))} indicates that higher values are better. The best-performing method for each metric within each training strategy is highlighted in \textbf{bold}.}

\label{tab:overall_results}
\end{table*}

\section{Experiments and Result Analysis}

\textbf{Dataset}: In this study, we evaluate our method using a Chinese speech dataset, which presents unique challenges due to its complex pronunciation and tonal variations compared to other languages, such as English. We use the Baker dataset~\cite{BakerDataset2020}, which contains approximately 12 hours of speech recorded using professional instruments at a frequency of 48kHz. The dataset consists of 10k speech samples from a female Mandarin speaker. 

\textbf{Experimental setups}: The experiment is conducted using an NVIDIA RTX A5000 with a PyTorch implementation. All experimental settings closely follow those proposed in StyleSpeech~\cite{lou2024stylespeechparameterefficientfinetuning}. Specifically, we use 4k sentences for training and 1k sentences for testing. The batch size is set to 64, and the model is trained for 300 epochs. The number of diffusion steps, $T$, is set to 50. To further validate our method, we also train our model on a larger dataset consisting of 9k training sentences and 1k testing sentences. An ablation study on the effect of the duration target $l$ was conducted to evaluate the impact of the duration adaptor on the output speech. In this study, phoneme samples adapted with the ground truth duration target are labelled as $(w/l)$, while those adapted using the adaptor-predicted duration are labelled as $(w/o l)$. Our source code will be released upon acceptance.

\textbf{Metrics}: We employ Word Error Rate~(WER), Mel Cepstral Distortion~(MCD)~\cite{kubichek1993mel}, and Perceptual Evaluation of Speech Quality~(PESQ)~\cite{rix2001perceptual}, to evaluate model's performance. For WER, we further evaluate the Phoneme-level WER (WER-P) and Style-level WER (WER-S). We assess the accuracy of synthesized speech using WER by first generating speech with a TTS system and then transcribing it through OpenAI's Whisper API~\cite{radford2023robust}.

\textbf{Result}: Table~\ref{tab:overall_results} presents results of our experiment. LatentSpeech shows significant improvements over FastSpeech and StyleSpeech. Specifically, it achieves a 25\% improvement in WER and a 24\% improvement in MCD compared to existing baseline models when trained on the 4k sentence dataset. These improvements further increase to 49.5\% and 26\%, respectively, when the model is trained with the larger 9k sentence dataset.

In terms of compactness, we compare the dimensions of our features in the latent space with mainstream approaches that use MelSpecs as intermediate features. For speech at 48kHz with a duration of 10 seconds, a MelSpec with dimensions $[80 \times 1873]$ (window length of 1024, hop length of 256, and 80 mel filters) is \textbf{20} times larger than our latent embedding of $[16 \times 469]$. This reduction means our method only requires \textbf{5\%} of the data dimensions needed by spectral representation. 

Figure~\ref{fig:audio_visual} presents embedding visualizations at different stages within the TTS system, including TTS Embedding $H_{TTS}$ (Figure~\ref{fig:tts_embed}), real and generated Latent Embeddings $Z$ (Figure~\ref{fig:latent_real} \& \ref{fig:latent_fake}), and MelSpecs for real and generated speeches (Figures~\ref{fig:real_mel} \& \ref{fig:fake_mel}). The MelSpec diagrams show a sparse data distribution, while the latent embeddings are more compact. This suggests that latent feature encoding utilizes the latent space more efficiently during speech encoding and decoding. Hence, it makes the encoding process more effective than traditional methods that encode speech to spectrograms using short Fast-Fourier transform. 

This significant reduction in data complexity benefits both the TTS encoder and the vocoder. With lower complexity, the TTS encoder requires fewer parameters and less computational load to map to the embeddings. It leads to a more accurate speech encoding process. Likewise, the vocoder generates more precise speech, as the compact latent embeddings preserve essential information without the interference caused by the sparsity observed in MelSpecs.

The results show that for the 4k sentence dataset, predictions with ground truth durations (w l) perform worse than those without $l$ (w/o l). Conversely, for the 9k sentence dataset, predictions with ground truth durations perform better. This difference arises from overfitting and the model's flexibility.
When using ground truth durations with a model trained on a smaller dataset, the limited data variety can cause the acoustic embeddings to overfit to specific durations seen during training. It reduce the model's flexibility to handle new durations for phoneme and style patterns. In contrast, using the model's predicted durations allows it to optimize acoustic features based on phoneme and style patterns, which leads to speech with higher clarity.
For larger datasets like 9k sentences, the model is exposed to a wider variety of durations and acoustic patterns. This increased data variety enhances the model's capacity to optimize acoustic patterns for different durations. Hence, (w l) proves more effective here because it closely matches how the speaker speaks. The performance difference between (w l) and (w/o l) for the larger dataset is subtle (less than 1\%). It indicates that both approaches are effective and the duration adaptor has successfully learned to predict accurate durations for each phoneme.

Regarding MCD, which measures the quality of speech generation in comparison with original speech. LatentSpeech (w l) achieves the best performance with MCD of 9.723 when trained on 4k sentences. It significantly outperform both FastSpeech and StyleSpeech. Further training with 9k sentences reduces the MCD to 9.498. However, it's worth noting that LatentSpeech (w/o l) has a higher MCD of 15.724. It suggests that duration label $l$ plays a crucial role in enhancing speech quality.
In terms of PESQ, which assesses the perceptual quality of the synthesized speech, LatentSpeech (w l) maintains competitive PESQ scores. It achieves a score of 1.055 for 4k sentences and 1.058 for 9k sentences. Interestingly, LatentSpeech (w/o l) achieves the highest PESQ score of 1.063 with 9k sentences. This indicates that while duration labels contribute to a lower MCD, they may not always improve perceptual quality, as seen in certain configurations where PESQ scores are higher without them.

\section{Conclusion}
In conclusion, we propose LatentSpeech, a new TTS framework that uses latent embeddings that reduce intermediate representation dimension to 5\% of mainstream approaches.
By incorporating a latent diffusion model, LatentSpeech refines speech in latent space for more accurate and natural output. Extensive experiments demonstrate that LatentSpeech achieves a 25\% improvement in WER and a 24\% improvement in MCD compared to existing models, with further improvements to 49.5\% and 26\% when trained with more data.
\newpage
\bibliographystyle{IEEEbib}
\bibliography{reference}

\begin{thebibliography}{10}

\bibitem{goodfellow2020generative}
Ian Goodfellow, Jean Pouget-Abadie, Mehdi Mirza, Bing Xu, David Warde-Farley, Sherjil Ozair, Aaron Courville, and Yoshua Bengio,
\newblock ``Generative adversarial networks,''
\newblock {\em Communications of the ACM}, vol. 63, no. 11, pp. 139--144, 2020.

\bibitem{kingma2013auto}
Diederik~P Kingma and Max Welling,
\newblock ``Auto-encoding variational bayes,''
\newblock {\em arXiv preprint arXiv:1312.6114}, 2013.

\bibitem{rombach2022high}
Robin Rombach, Andreas Blattmann, Dominik Lorenz, Patrick Esser, and Bj{\"o}rn Ommer,
\newblock ``High-resolution image synthesis with latent diffusion models,''
\newblock in {\em Proceedings of the IEEE/CVF conference on computer vision and pattern recognition}, 2022, pp. 10684--10695.

\bibitem{ramesh2022hierarchical}
Aditya Ramesh, Prafulla Dhariwal, Alex Nichol, Casey Chu, and Mark Chen,
\newblock ``Hierarchical text-conditional image generation with clip latents,''
\newblock {\em arXiv preprint arXiv:2204.06125}, vol. 1, no. 2, pp. 3, 2022.

\bibitem{ho2022imagen}
Jonathan Ho, William Chan, Chitwan Saharia, Jay Whang, Ruiqi Gao, Alexey Gritsenko, Diederik~P Kingma, Ben Poole, Mohammad Norouzi, David~J Fleet, et~al.,
\newblock ``Imagen video: High definition video generation with diffusion models,''
\newblock {\em arXiv preprint arXiv:2210.02303}, 2022.

\bibitem{wang2017tacotron}
Yuxuan Wang, RJ~Skerry-Ryan, Daisy Stanton, Yonghui Wu, Ron~J Weiss, Navdeep Jaitly, Zongheng Yang, Ying Xiao, Zhifeng Chen, Samy Bengio, et~al.,
\newblock ``Tacotron: Towards end-to-end speech synthesis,''
\newblock {\em arXiv preprint arXiv:1703.10135}, 2017.

\bibitem{ren2019fastspeech}
Yi~Ren, Yangjun Ruan, Xu~Tan, Tao Qin, Sheng Zhao, Zhou Zhao, and Tie-Yan Liu,
\newblock ``Fastspeech: Fast, robust and controllable text to speech,''
\newblock {\em Advances in neural information processing systems}, vol. 32, 2019.

\bibitem{lou2024stylespeechparameterefficientfinetuning}
Haowei Lou, Helen Paik, Wen Hu, and Lina Yao,
\newblock ``Stylespeech: Parameter-efficient fine tuning for pre-trained controllable text-to-speech,'' 2024.

\bibitem{zhang2023survey}
Chenshuang Zhang, Chaoning Zhang, Sheng Zheng, Mengchun Zhang, Maryam Qamar, Sung-Ho Bae, and In~So Kweon,
\newblock ``A survey on audio diffusion models: Text to speech synthesis and enhancement in generative ai,''
\newblock {\em arXiv preprint arXiv:2303.13336}, 2023.

\bibitem{liu2023diffvoice}
Zhijun Liu, Yiwei Guo, and Kai Yu,
\newblock ``Diffvoice: Text-to-speech with latent diffusion,''
\newblock in {\em ICASSP 2023-2023 IEEE International Conference on Acoustics, Speech and Signal Processing (ICASSP)}. IEEE, 2023, pp. 1--5.

\bibitem{ren2020fastspeech}
Yi~Ren, Chenxu Hu, Xu~Tan, Tao Qin, Sheng Zhao, Zhou Zhao, and Tie-Yan Liu,
\newblock ``Fastspeech 2: Fast and high-quality end-to-end text to speech,''
\newblock {\em arXiv preprint arXiv:2006.04558}, 2020.

\bibitem{caillon2021rave}
Antoine Caillon and Philippe Esling,
\newblock ``Rave: A variational autoencoder for fast and high-quality neural audio synthesis,''
\newblock {\em arXiv preprint arXiv:2111.05011}, 2021.

\bibitem{nguyen1994near}
Truong~Q Nguyen,
\newblock ``Near-perfect-reconstruction pseudo-qmf banks,''
\newblock {\em IEEE Transactions on signal processing}, vol. 42, no. 1, pp. 65--76, 1994.

\bibitem{engel2020ddsp}
Jesse Engel, Lamtharn Hantrakul, Chenjie Gu, and Adam Roberts,
\newblock ``Ddsp: Differentiable digital signal processing,''
\newblock {\em arXiv preprint arXiv:2001.04643}, 2020.

\bibitem{vaswani2017attention}
Ashish Vaswani, Noam Shazeer, Niki Parmar, Jakob Uszkoreit, Llion Jones, Aidan~N Gomez, {\L}ukasz Kaiser, and Illia Polosukhin,
\newblock ``Attention is all you need,''
\newblock {\em Advances in neural information processing systems}, vol. 30, 2017.

\bibitem{kong2020diffwave}
Zhifeng Kong, Wei Ping, Jiaji Huang, Kexin Zhao, and Bryan Catanzaro,
\newblock ``Diffwave: A versatile diffusion model for audio synthesis,''
\newblock {\em arXiv preprint arXiv:2009.09761}, 2020.

\bibitem{ho2020denoising}
Jonathan Ho, Ajay Jain, and Pieter Abbeel,
\newblock ``Denoising diffusion probabilistic models,''
\newblock {\em Advances in neural information processing systems}, vol. 33, pp. 6840--6851, 2020.

\bibitem{BakerDataset2020}
Databaker,
\newblock ``Chinese mandarin female corpus,'' \url{https://en.data-baker.com/datasets/freeDatasets/}, 2020,
\newblock Accessed: 2023-04-20.

\bibitem{kubichek1993mel}
Robert Kubichek,
\newblock ``Mel-cepstral distance measure for objective speech quality assessment,''
\newblock in {\em Proceedings of IEEE pacific rim conference on communications computers and signal processing}. IEEE, 1993, vol.~1, pp. 125--128.

\bibitem{rix2001perceptual}
Antony~W Rix, John~G Beerends, Michael~P Hollier, and Andries~P Hekstra,
\newblock ``Perceptual evaluation of speech quality (pesq)-a new method for speech quality assessment of telephone networks and codecs,''
\newblock in {\em 2001 IEEE international conference on acoustics, speech, and signal processing. Proceedings (Cat. No. 01CH37221)}. IEEE, 2001, vol.~2, pp. 749--752.

\bibitem{radford2023robust}
Alec Radford, Jong~Wook Kim, Tao Xu, Greg Brockman, Christine McLeavey, and Ilya Sutskever,
\newblock ``Robust speech recognition via large-scale weak supervision,''
\newblock in {\em International Conference on Machine Learning}. PMLR, 2023, pp. 28492--28518.

\end{thebibliography}

\end{document}